\documentclass[final,3p,times,twocolumn]{elsarticle}
 
\usepackage{amssymb}
\usepackage{amsmath,amssymb,exscale}
\usepackage{graphicx}
\usepackage{epsfig}
\usepackage{multicol}
\usepackage{color}
\usepackage{xcolor}
\usepackage{hyperref}
\usepackage{mathrsfs}
\usepackage{hyperref}
\hypersetup{colorlinks,bookmarksopen,bookmarksnumbered,citecolor=rossos,
linkcolor=black,pdfstartview=FitH,urlcolor=blus}
\usepackage{amsfonts}
\usepackage{slashed}
\usepackage{blindtext}
 \usepackage{fancyhdr}
\usepackage{hyperref}

\allowdisplaybreaks

\definecolor{blus}{cmyk}{1,1,0,0.}
\definecolor{verdes}{cmyk}{0.99,0,0.99,0.02}
\definecolor{rossos}{cmyk}{0,1,1,0.55}
\definecolor{greeny}{cmyk}{0.99,0,0.59,0.98}
\definecolor{redy}{cmyk}{0,1,1,0.40}

\def\bp{\bar M_{\rm Pl}}
\def\Lag{\mathscr{L}}

\def\be{\begin{equation}}
\def\ee{\end{equation}}
\def\bea{\begin{eqnarray}}
\def\eea{\end{eqnarray}}
\def\ba{\begin{array} }
\def\ea{\end{array}}
\def\bac{\begin{array} {c}}
\def\bacc{\begin{array} {cc}}
\def\baccc{\begin{array} {ccc}}

\def\psl{\hbox{\hbox{${p}$}}\kern-1.9mm{\hbox{${/}$}}}
\def\dsl{\hbox{\hbox{${\partial}$}}\kern-1.7mm{\hbox{${/}$}}}
\def\Dsl{\hbox{\hbox{${D}$}}\kern-2.1mm{\hbox{${/}$}}}
\definecolor{red}{rgb}{1,0,0}

\def\hhref#1{\href{http://arxiv.org/abs/#1}{arXiv:#1}} 
\hypersetup{colorlinks=true, linkcolor=red}

\journal{the arXiv}

\begin{document}

\begin{frontmatter}

\title{\vspace{-2cm}\huge {\color{redy}Critical Higgs inflation in a Viable Motivated Model}}

\author{\vspace{1cm}{\large {\bf Alberto Salvio}}}

\address{\normalsize \vspace{0.2cm}CERN, Theoretical Physics Department, Geneva, Switzerland  \\
and\\
Dipartimento di Fisica, Universit\`a di Roma and INFN Tor Vergata \\

\vspace{0.3cm}
 {\it {\small Report numbers: CERN-TH-2018-213}}
 \vspace{-1cm}
 }

\begin{abstract}
An extension of the Standard Model  with three right-handed neutrinos and a simple invisible axion model  can account for all experimentally confirmed signals of new physics (neutrino oscillations, dark matter and baryon asymmetry) in addition to solving the strong CP problem, stabilizing the electroweak  vacuum and satisfying all current observational bounds.  
We show that this model can also implement critical Higgs inflation, which corresponds to the frontier between stability and metastability of the electroweak vacuum. This leads to a value of the non-minimal coupling between the Higgs and the Ricci scalar that is much lower than the one usually quoted in Higgs inflation away from criticality. Then, an advantage is that the scale of perturbative unitarity breaking on flat spacetime can be very close to the Planck mass, where anyhow new physics is required. The higher dimensional operators are under control in this inflationary setup. The dependence of the cutoff on the Higgs background is also taken into account as appropriate when the Higgs is identified with the inflaton. Furthermore, critical Higgs inflation enjoys a robust inflationary attractor that makes it an appealing setup for the early universe.
In the proposed model, unlike in  the Standard Model, critical Higgs inflation can be realized without any tension with the observed quantities, such as the top mass and the strong coupling. 
\end{abstract}


\end{frontmatter}
 
 \begingroup
\hypersetup{linkcolor=blus}
\tableofcontents
\endgroup



\newpage

\section{Introduction}\label{introduction}
 
  It has been shown that extending the Standard Model (SM) with three right-handed neutrinos (with a generic flavor structure) and with the extra fields required by a simple invisible axion model can solve the observational problems of the SM (neutrino oscillations, dark matter (DM) and baryon asymmetry) and eliminate a number of unsatisfactory aspects of the SM~\cite{Salvio:2015cja,Ballesteros:2016euj,Ballesteros:2016xej}. These include the strong CP problem and the metastability\footnote{The electroweak vacuum is metastable when unstable, but with a lifetime larger than the age of the universe.} of the electroweak (EW) vacuum. 

 The invisible axion model considered in~\cite{Salvio:2015cja} and later further studied in~\cite{Ballesteros:2016euj,Ballesteros:2016xej} is perhaps the simplest  model  of this sort (originally proposed by Kim, Shifman, Vainshtein and Zakharov (KSVZ)~\cite{Kim:1979if}), in which one introduces the following extra fields 
\begin{itemize}
\item {\bf An extra Dirac fermion.} This Dirac fermion $Q$ consists of a pair of two-component Weyl fermions $q_1$ and $q_2$ in the following representation of the SM gauge group $G_{\rm SM}\equiv {\rm SU(3)_c\times SU(2)_{\it L}\times U(1)_{\it Y}}$ 
\be q_1 \sim (3,1)_0, \qquad q_2 \sim (\bar{3}, 1)_0.\ee
The $q_i$ are charged under a spontaneously broken and anomalous axial U(1) symmetry present in any axion model, the Peccei-Quinn (PQ) symmetry~\cite{Peccei:1977hh}.
\item {\bf An extra complex scalar.} This scalar $A$ is charged under the PQ symmetry  and neutral under $G_{\rm SM}$. The PQ charge of $A$ is   
twice as large as that of the $q_i$, such that a Yukawa coupling between $A$ and $Q$ can be non-zero.
\end{itemize}

Given that  a single quark flavor carrying a non-vanishing PQ charge is present, the model avoids the domain wall problem~\cite{Sikivie:1982qv}, as discussed in Ref.~\cite{Barr:1982uj}.

 The above-mentioned extra fields can render the EW vacuum absolutely stable and,  therefore,  one can identify the inflaton with the Higgs~\cite{Salvio:2015cja}. Indeed, the condition to have successful Higgs inflation (HI)~\cite{CervantesCota:1995tz,Bezrukov:2007ep} turns out to be very similar to that of vacuum stability~\cite{Bezrukov:2009db,Bezrukov:2009-2,Salvio-inf}. However, it was pointed out that the original implementation of HI proposed in~\cite{Bezrukov:2007ep} leads to the breaking of perturbative unitarity well below the Planck scale when a perturbative expansion around the {\it flat spacetime} is performed~\cite{crit}. This is due to the fact that the HI of~\cite{Bezrukov:2007ep} requires a large non-minimal coupling $\xi_H$ between the Higgs and the Ricci scalar and, consequently, a new scale $\bp/\xi_H$ is generated~\cite{crit}, where $\bp$ is the reduced Planck mass. Although this does not necessarily invalidate the HI of~\cite{Bezrukov:2007ep} as the SM can enter strong coupling when collisions occur at energies $\bp/\xi_H$ and on {\it flat spacetime}\footnote{Indeed, the spacetime is not flat during inflation and, therefore, it is still possible that during this phase of the early universe perturbation theory is reliable.}, another possible interpretation of the breaking of perturbative unitarity is the onset of new physics, which could change the inflationary predictions. For this reason Refs.~\cite{Ballesteros:2016euj,Ballesteros:2016xej} proposed to identify the inflaton with $|A|$ or a combination of $|A|$ and the Higgs.
Furthermore, in~\cite{Salvio:2015kka} it was shown\footnote{See also~\cite{Calmet:2016fsr} for a subsequent discussion.} that, unless some parameters are strongly fine-tuned,  a large $\xi_H$ can generate higher order operators in the quantum effective action, which can change the inflationary predictions. 

 However, the large value of $\xi_H$ used in~\cite{Bezrukov:2007ep} can be drastically reduced by taking quantum corrections into account~\cite{Bezrukov:2009-2,Salvio-inf,Allison:2013uaa}. The minimal value of $\xi_H$ is achieved by living very close to the frontier between metastability and stability of the EW vacuum, implementing the so-called critical Higgs inflation (CHI)~\cite{Hamada:2014iga,Bezrukov:2014bra,Hamada:2014wna}. In this case, the scale of perturbative  unitarity breaking can be essentially  identified with the scale at which Einstein's theory of gravity breaks down. Of course, at those Planckian energies, some new physics is anyhow required to UV-complete gravity. Moreover, in CHI higher dimensional operators do not significantly change the predictions and the inflationary dynamics enjoys a robust attractor~\cite{Salvio:2017oyf}. The latter property is very important: if it were not satisfied one would have  to fine-tune the initial conditions of the inflaton, and this  would make the whole idea of inflation less attractive. The way inflation takes place in CHI is substantially different from the original HI of~\cite{Bezrukov:2007ep} as the potential in the critical case features a quasi-inflection point\footnote{See Refs.~\cite{Hamada:2014iga,Bezrukov:2014bra,Lloyd-Stubbs:2018ouj} for a previous study of inflection points in the SM and in the KSVZ model.}. Reheating can also be successfully implemented in HI~\cite{Bezrukov:2008ut} (both in the critical and non-critical versions) because the Higgs has sizable couplings to other SM particles; this leads to a high reheating temperature, $T_{\rm RH}\gtrsim 10^{13}$~GeV.  Another reason for considering CHI in this model is the fact that, after a careful analysis,  Refs.~\cite{Ballesteros:2016euj,Ballesteros:2016xej} found that the pure $|A|$-inflation is not viable and they eventually proposed a multifield inflation in which both the Higgs and $|A|$ vary along the inflationary path; (critical) Higgs inflation, on the other hand, can offer the possibility to achieve the simpler single-field option.

 Given these advantages of CHI we here explore whether this version of HI can be implemented in the well-motivated SM extension that includes the KSVZ axion model and three right-handed neutrinos~\cite{Salvio:2015cja}. Moreover, we investigate whether the cutoff of the theory is always bigger than the typical energies taking into account the background Higgs field, as appropriate when the Higgs is identified with the inflaton and, therefore, has  a large field value during inflation.  We here focus on the original model of~\cite{Salvio:2015cja} because the action of~\cite{Salvio:2015cja} is simpler than that of \cite{Ballesteros:2016euj,Ballesteros:2016xej}  thanks to a different choice of symmetries (see the next section).

 The article is organized as follows. In the next section we give further details of the model, which will give us the opportunity to introduce the notation. In Sec.~\ref{Observational bounds} we discuss the current observational bounds updating the analysis of~\cite{Salvio:2015cja} with new experimental and observational results. The renormalization group equations needed to compute the effective potential are presented in Sec.~\ref{Renormalization group equations} including those of the non-minimal couplings between the scalars and gravity and 2-loop extensions. Sec.~\ref{Stability analysis} is instead devoted to the analysis of the stability of the EW vacuum, which is more involved than in the SM due to the presence of an extra scalar, $A$. The actual analysis of inflation is only performed in Sec.~\ref{crit} because the new insight provided by the previous sections is necessary for a detailed inflationary analysis. Finally, in Sec.~\ref{Validity of the effective theory}   the cutoff of the theory is investigated taking into account the Higgs background in CHI. Our conclusions are presented in Sec.~\ref{Conclusions}.

\vspace{-0.1cm}

\section{The model}\label{model}

 Let us now give a detailed description of the model. Here we consider the SM plus three right-handed neutrinos $N_i$ and the extra fields of the first viable invisible axion model (the KSVZ model~\cite{Kim:1979if})~\cite{Salvio:2015cja}. The gauge group of the model is the SM group $G_{\rm SM}$.

 The Lagrangian is given by
 \be \mathscr{L} = \Lag_{\rm SM}+\Lag_{N}+  \Lag_{\rm axion}+ \Lag_{\rm gravity}.\label{full-lagrangian} \ee
 We define in turn the various terms in $ \mathscr{L}$ above.  $\Lag_{\rm SM}$ corresponds to the SM Lagrangian, while  $\Lag_{N}$ represents the part of the Lagrangian that depends on the $N_i$:
  \be  i\overline{N}_i \dsl N_i+ \left(\frac12 N_i M_{ij}N_j +  Y_{ij} L_iH N_j + {\rm h.  c.}\right). \ee
  $M_{ij}$ is the Majorana mass matrix of  $N_i$
and $Y_{ij}$ is the neutrino Yukawa coupling matrix governing the interaction with the SM Higgs doublet $H$ and the standard  lepton doublets $L_i$. Notice that the matrix $M$ can be taken symmetric without loss of generality, but generically it has  complex elements. However, we assume it to be diagonal and real without loss of generality thanks to the complex Autonne-Takagi factorization. 
 So
  $$M=\mbox{diag}(M_1, M_2, M_3),$$ where the $M_i$  ($i=1,2,3$) are  the Majorana masses of the three right-handed neutrinos.

  $\Lag_{\rm axion}$ gives the additional terms in the Lagrangian due to the KSVZ model:
  \be \mathscr{L}_{\rm axion} = i\sum_{j=1}^2\overline{q}_j \Dsl \, q_j +|\partial A|^2  -(y q_2A q_1 +h.c.)-\Delta V(H,A)\nonumber \ee
The full classical potential is
\be  V(H,A)=  \lambda_H(|H|^2-v^2)^2+\Delta V(H,A), \ee
where 
\be \Delta V(H,A) \equiv \lambda_A(|A|^2-  f_a^2)^2 + \lambda_{HA} (|H|^2-v^2)( |A|^2-f_a^2),\nonumber \ee
 and $v$ and $f_a$ are real and positive parameters, which can be interpreted as the EW and PQ breaking scales, respectively.
  The Yukawa coupling  $y$ of $Q$ is chosen real without loss of generality.
The quartic couplings $\lambda_H$, $\lambda_A$ and $\lambda_{HA}$ have to satisfy some bounds to ensure the stability of the EW vacuum, as we will see in Sec.~\ref{Stability analysis}.

The PQ symmetry acts on $q_1$, $q_2$ and $A$ as follows
\be q_1\rightarrow e^{i\alpha/2}q_1, \quad q_2\rightarrow e^{i\alpha/2}q_2, \quad A\rightarrow e^{-i\alpha}A, \ee
where $\alpha$ is an arbitrary real parameter. This symmetry forbids  a tree level mass term  $M_q q_1 q_2  +h.c.$. The SM fields and the right-handed neutrinos are not charged under U(1)$_{\rm PQ}$. The model has the accidental symmetry
\be q_1\rightarrow  - q_1 ,\quad q_2\rightarrow  q_2,\quad A\rightarrow -A .
\ee 

Finally, $ \Lag_{\rm gravity}$ are the terms in the Lagrangian that include the pure gravitational part and the possible non-minimal couplings between gravity and the other fields:
 \be \mathscr{L}_{\rm gravity} = -\left(\frac{\bp^2}{2} +\xi_H (|H|^2-v^2) + \xi_A (|A|^2-f_a^2)\right)\mathscr{R} -\Lambda,
 \label{gravity-Lag} \ee
where $\bar{M}_{\rm Pl}\simeq 2.4\times 10^{18}\,$GeV is the reduced Planck mass, $\mathscr{R}$ is the Ricci scalar,
 $\xi_H$ and $\xi_A$ are the non minimal couplings of the Higgs and the new scalar to gravity and $\Lambda$ is the cosmological constant, which is introduced to account for dark energy.

The EW symmetry breaking is triggered by the vacuum expectation value (VEV) $v\simeq 174\,$GeV of the neutral component $H_0$ of $H$ (while all the other components of $H$ have a vanishing VEV). After that the neutrinos acquire a Dirac mass matrix
$ m_D = v Y,$
  which can  be parameterized in terms of  column vectors $m_{Di}$ ($i=1,2,3$):
\be m_D =\left(\begin{array}{ccc}\hspace{-0.1cm}m_{D1}\,, & \hspace{-0.2cm}m_{D2}\, ,  & \hspace{-0.2cm} m_{D3}\hspace{-0.1cm}
\end{array}\right). \label{Dirac-mass}\ee
   Integrating out the heavy neutrinos $N_i$, one then obtains  the following light neutrino Majorana mass matrix  
\be m_\nu =  \frac{m_{D1} m_{D1}^T}{M_1} + \frac{m_{D2} m_{D2}^T}{M_2} + \frac{m_{D3} m_{D3}^T}{M_3} . \label{see-saw} \ee 
  By means of a unitary (Autonne-Takagi) redefinition of the left-handed neutrinos we can diagonalize $m_\nu$ obtaining the mass eigenvalues $m_1, m_2$ and $m_3$ (the left-handed neutrino Majorana masses). Calling $U_\nu$ the unitary matrix that implements such transformation (a.k.a. the Pontecorvo-Maki-Nakagawa-Sakata (PMNS) matrix) i.e. $U^T_\nu m_\nu U_\nu= $ diag$(m_1, m_2, m_3)$, we can parameterize $U_\nu= V_\nu P_{12}$, where
\be 
\small V_\nu=\left(\begin{array}{ccc}
c_{12} c_{13} & s_{12} c_{13} & s_{13} e^{-i \delta}  \\ -s_{12}c_{23}-c_{12}s_{13}s_{23}e^{i\delta} & c_{12}c_{23}-s_{12}s_{13}s_{23}e^{i\delta}  & c_{13}s_{23} \\ 
s_{12}s_{23}-c_{12} s_{13} c_{23}e^{i\delta}  & -c_{12}s_{23}-s_{12}s_{13} c_{23} e^{i\delta}& c_{13}c_{23}
\end{array}\right),
\nonumber\ee
with $s_{ij}\equiv \sin(\theta_{ij})$, $c_{ij}\equiv \cos(\theta_{ij})$; $\theta_{ij}$ are the neutrino mixing angles and $P_{12}$ is a diagonal matrix that  contains two extra phases: 
\be P_{12}=\left(\begin{array}{ccc}
e^{i \beta_1} & 0  & 0  \\  0 & e^{i\beta_2} & 0\\ 
0  & 0 & 1
\end{array}\right).   \ee
Even in the most general case of three right-handed neutrinos, it is possible to express $Y$ in terms of low-energy observables, the heavy masses  $M_1$, $M_2$ and $M_3$ and extra parameters~\cite{Casas:2001sr}: 
\be Y= \frac{U_\nu^* D_{\sqrt{m}}\, R \, D_{\sqrt{M}} }{v}, \label{Casas-Ibarra}\ee
where 
\bea D_{\sqrt{m}}&\equiv& \mbox{diag}(\sqrt{m_1},\sqrt{m_2},\sqrt{m_3}),\nonumber \\ D_{\sqrt{M}}&\equiv& \mbox{diag}(\sqrt{M_1},\sqrt{M_2}, \sqrt{M_3}) \nonumber\eea
and $R$ is a generic complex orthogonal matrix (that contains the extra parameters). This is useful for us because the observational constraints are not directly on $Y$, but they are rather on the low-energy quantities $m_i$, $U_\nu$ and on $M_i$ (see section \ref{Observational bounds}).  One can show that the simplest and realistic case of  two right-handed neutrinos \cite{Ibarra:2005qi} below $M_{\rm Pl}$ can be recovered by setting $m_1=0$ and 
\be R = \left(\begin{array}{ccc}
0 & 0  & 1\\ \cos z & -\sin z & 0\\ 
\xi \sin z & \xi \cos z & 0
\end{array}\right),\nonumber\ee 
where $z$ is a complex parameter and $\xi=\pm 1$.

The PQ symmetry is spontaneously broken by $f_a\equiv \langle A\rangle$, leading to the following squared mass of  $Q$:  $$M_q^2 = y^2 f_a^2.$$ Moreover, $A$ contains a (classically) massless particle, the axion, and  a massive particle with squared mass
\be M_A^2 = f_a^2\left(4\lambda_A +\mathcal{O}\left(\frac{v^2}{f_a^2}\right)\right).  \label{MA1} \ee
Given the lower bound on $f_a$ that will be reviewed in Sec.~\ref{Observational bounds}, the corrections $\mathcal{O}\left(v^2/f_a^2\right)$ are very small and will be neglected in the following.

When the scalars are set to their VEV, $\mathscr{L}_{\rm gravity}$
 reduces to the standard pure Einstein-Hilbert action (with a cosmological constant), which is why we added the extra terms proportional to $v^2$ and $f_a^2$ in Eq.~(\ref{gravity-Lag}).

  \section{Observational bounds}\label{Observational bounds}

    \subsection*{Neutrino masses and oscillations}
    
 As far as the neutrino masses $m_i$ ($i=1,2,3$) are concerned, we have several data from oscillation and non-oscillation experiments. For example, Refs.~\cite{deSalas:2017kay,Capozzi:2018ubv} presented some of the most recent determinations of  $\Delta m^2_{21}$, $\Delta m^2_{3l}$ (where $\Delta m_{ij}^2 \equiv m_i^2-m_j^2$ and $ \Delta m^2_{3l} \equiv  \Delta m^2_{31} $ for normal ordering and $ \Delta m^2_{3l} \equiv  - \Delta m^2_{32} $ for inverted ordering), the mixing angles $\theta_{ij}$ and $\delta$. 
 
 Here we take the central values reported in~\cite{Capozzi:2018ubv} for normal ordering. Indeed,  normal ordering is currently favored over inverted ordering. 
Currently no significant constraints are known for $\beta_1$ and $\beta_2$; thus we will set these parameters to zero for simplicity from now on.

     \subsection*{Baryon asymmetry}
     
 Successful leptogenesis\footnote{Neutrino oscillations offer another mechanism to generate the baryon asymmetry through a different version of leptogenesis~\cite{Akhmedov:1998qx}. We do not consider this possibility in the numerical examples below, but it can be easily implemented in this model.}   \cite{fuk}
occurs if neutrinos are lighter than 0.15 eV and the lightest right-handed neutrino mass $M_l$ fulfills ~\cite{Davidson:2002qv}  
\be  M_l \gtrsim 1.7 	\times 10^{7}\, \mbox{GeV}.\label{leptobound} \ee
In order to be conservative we have reported the weakest bound, but depending on assumptions  one can have stronger bounds\footnote{For example if the initial abundance of right-handed neutrinos at $T\gg M_i$ is zero then  $M_l \gtrsim 2.4\times 10^{9}$~GeV ~\cite{Davidson:2002qv}.}.  

\subsection*{Constraints on the axion sector}

 In order not to overproduce DM through the misalignment mechanism~\cite{axionDM}  and to elude axion detection one obtains respectively an upper and lower bound (see e.g.~\cite{Kawasaki:2013ae}
 and~\cite{Raffelt:1999tx}, respectively) 
 on the order of magnitude of the scale of PQ symmetry breaking $f_a$:
\be 10^8 \, \mbox{GeV} \lesssim f_a \lesssim   10^{12} \, \mbox{GeV}. \label{bound-f}\ee
The upper bound is obtained by requiring that the axion field takes a value of order $f_a$ at early times, which is what we  expect but is not necessarily the case; also the precise value of the lower bound is model dependent. Therefore,  (\ref{bound-f}) should not be interpreted as sharp bounds, but it certainly gives a plausible range of $f_a$. The window in~(\ref{bound-f}) also allows us to neglect PQ symmetry breaking effects due to gravity: non-perturbative gravitational effects can violate PQ invariance, but lead to a sizable correction only for $f_a\gtrsim 10^{16}$~GeV (see Ref.~\cite{Hebecker:2018ofv} for a recent review).

In addition to contributing to DM, the axion also necessarily leads to dark radiation because it is also thermally produced~\cite{Masso, Graf:2010tv, Salvio:2013iaa}. This population of hot axions contributes to the effective number of relativistic species, but the size of this contribution is currently well within the observational bounds although, interestingly enough, within the reach of future observations in some models~\cite{Salvio:2013iaa,Ferreira:2018vjj} .

In the case of the KSVZ-based model considered here a more precise version of    the lower bound in (\ref{bound-f}) is $f_a\gtrsim 4 \times 10^8$\, GeV~\cite{Raffelt:1999tx}. In any case bounds on $f_a$  can only constrain the ratio $M_A/\sqrt\lambda_A$ as it is clear from (\ref{MA1}). When  $M_q\gg v$ and $M_A\gg v$ the EW constraints are fulfilled. The size of $y$ is also very mildly constrained: we have  a lower bound from the lack of observation, which is not more stringent than $M_q \gtrsim 1$ TeV (indeed one has to take into account that the extra quark $Q$ is {\it not} charged under the electroweak part of the SM gauge group).   
 Moreover, in this model the bounds on $f_a$, which allows the axion to account for the whole DM, is~\cite{Ballesteros:2016euj}
\be 2 \times 10^{10} \, {\rm GeV}  \lesssim f_a \lesssim  0.9\times  10^{11} \, \mbox{GeV}. \label{bound-f-DM}\ee
  
  \subsection*{Constraints on SM parameters}
  
  Finally, in order to have ``initial conditions" for the renormalization group equations (RGEs)\footnote{The RGEs of the model will be discussed in Sec.~\ref{Renormalization group equations}.}, we also have to fix the values of the relevant SM couplings at the EW scale, say at the top mass $M_t\simeq 172.5$~GeV~\cite{Castro:2017yxe,Pearson:2017jck}. We take the values computed in~\cite{Buttazzo:2013uya}, which expresses these quantities in terms of $M_t$, the Higgs mass $M_h\simeq 125.09$~GeV~\cite{Aad:2015zhl}, the strong fine-structure constant renormalized at the Z mass, $\alpha_s(M_Z) \simeq 0.1184$~\cite{Bethke:2012jm} and $M_W\simeq 80.379$~GeV~\cite{Particle data group} (see the quoted literature for the uncertainties on these quantities).

\subsection*{Inflation}

In 2018 Planck released new results for inflationary observables~\cite{Planck2018}, which are relevant for our purposes. 

For example, for the scalar spectral index $n_s$ and the tensor-to-scalar ratio $r$ one has now
\be n_s= 0.9649 \pm 0.0042 \, \,  (68\% {\rm CL}), \quad r< 0.064 \, \,  (95\% {\rm CL}), \label{nsr}\ee
while for the curvature power spectrum $P_R(q)$ (at horizon exit\footnote{We  use a standard notation:   $a$ is the cosmological scale factor, $H\equiv \dot a/a$ and a dot denotes the derivative with respect to (cosmic) time, $t$.} $q=aH$) 
\be P_R =(2.10 \pm 0.03) 10^{-9} . \label{PRobserved}\ee 
These constraints are particularly important for us as the main goal of the article is to study whether CHI is viable.

\section{Renormalization group equations}\label{Renormalization group equations}

Given that we want to obtain the predictions of this model at energies much above the EW scale, we need the complete set of RGEs.  We adopt the $\overline{\rm MS}$ renormalization scheme  to define the renormalized couplings. Moreover, for a generic coupling $g$ we write the RGEs as
\be \frac{dg}{d\tau}= \beta_{g},\ee
where $d/d\tau\equiv \bar{\mu}^2\, d/d\bar{\mu}^2$ and $\bar{\mu}$ is the $\overline{\rm MS}$ renormalization energy scale. The $\beta$-functions  $\beta_{g}$ can also be expanded in loops: 
\be  \beta_{g} =  \frac{\beta_{g}^{(1)}}{(4\pi)^2}+ \frac{\beta_{g}^{(2)}}{(4\pi)^4}+ ... \, ,\ee 
where $ \beta_{g}^{(n)}/(4\pi)^{2n}$   is the $n$-loop contribution. 

We start from energies much above $M_A$, $M_q$ and $M_{ij}$. In this case, the 1-loop RGEs  are (see \cite{Pirogov:1998tj,EliasMiro:2011aa,EliasMiro:2012ay,Salvio:2014soa,Salvio:2015cja} for previous determinations of some terms in some of these  RGEs)
%
\bea  \beta_{g_1^2}^{(1)}& =&    \frac{41g_1^4}{10}, \qquad   \beta_{g_2^2}^{(1)} =- \frac{19g_2^4}{6},\qquad\beta_{g_3^2}^{(1)}  = -\frac{19 g_3^4}{3},\nonumber\\   \beta_{y_t^2}^{(1)}  & =& y_t^2\left(\frac92 y_t^2-8g_3^2-\frac{9g_2^2}{4}-\frac{17g_1^2}{20} + {\rm Tr}(Y^\dagger Y )\right),\nonumber\\ 
  \beta_{\lambda_H}^{(1)} & =&\left(12\lambda_H+6y_t^2-\frac{9g_1^2}{10}-\frac{9g_2^2}{2}+2\, {\rm Tr}(Y^\dagger Y)\right)\lambda_H \nonumber\\ &&\hspace{-0.7cm}-\, 3y_t^4 +\frac{9 g_2^4}{16}+\frac{27 g_1^4}{400}+\frac{9 g_2^2 g_1^2}{40}+\frac{\lambda_{HA}^2}{2} - {\rm Tr}((Y^\dagger Y)^2), \nonumber\\ 
 \beta_{\lambda_{HA}}^{(1)} & =& \left(3y_t^2-\frac{9g_1^2}{20}-\frac{9g_2^2}{4}+6\lambda_H \right) \lambda_{HA}\nonumber \\ && +\left(4\lambda_A +\, {\rm Tr}(Y^\dagger Y ) + 3y^2 \right) \lambda_{HA}+2 \lambda_{HA}^2, \nonumber\\ 
 \beta_{\lambda_A}^{(1)} & =& \lambda_{HA}^2+10\lambda_A^2+6y^2 \lambda_A- 3 y^4,\nonumber\\ 
   \beta_{Y}^{(1)} & =&Y  \left[\frac32 y_t^2-\frac{9}{40} g_1^2-\frac98 g_2^2+\frac34 Y^\dagger Y+\frac12 {\rm Tr}(Y^\dagger Y )\right],\nonumber \\ 
 \beta_{y^2}^{(1)} & =&y^2(4y^2-8 g_3^2),\nonumber \\ 
  \beta_{\xi_H}^{(1)} &=& (1+6\xi_H)\left(\frac{y_t^2}{2}+\frac{{\rm Tr}(Y^\dagger Y )}6 -\frac{3g_2^2}8  - \frac{3 g_1^2}{40}+\lambda_H\right)\nonumber \\ &&-\frac{\lambda_{HA}}{6}(1+6\xi_A), \nonumber \\
 \beta_{\xi_A}^{(1)}  &=& (1+6\xi_A) \left(\frac{y^2}{2}+\frac{2}{3}\lambda_A\right)
-\frac{\lambda_{HA}}{3} (1+6\xi_H), \nonumber
\eea
where  $g_3$,  $g_2$ and  $g_1=\sqrt{5/3}g_Y$ are the gauge couplings of ${\rm SU(3)}_c $, ${\rm SU(2)_{\it L}}$ and   ${\rm U(1)_{\it Y}}$, respectively, and $y_t$ is the top Yukawa coupling.  In addition to the $\beta$-functions presented in~\cite{Salvio:2015cja} we have added here the RGEs for the non-minimal couplings $\xi_H$ and $\xi_A$, which, as we will see, play some role in inflation.

Since the SM couplings evolve in the full range from the EW to the Planck scale it is appropriate to use for them the 2-loop RGEs\footnote{In the absence of gravity the RGEs for a generic quantum field theory  were computed up to 2-loop order in~\cite{MV}.}, which we present explicitly here for the first time including the new physics contribution:
\bea 
 \beta_{g_1^2}^{(2)}&=&g_1^4 \left(\frac{199 g_1^2}{50}+\frac{27 g_2^2}{10}+\frac{44 g_3^2}{5}-\frac{17 y_t^2}{10}\right. \nonumber \\ &&\left.-\frac{3 }{10} {\rm Tr}(Y^\dagger Y)\right), \nonumber 
\\
 \beta_{g_2^2}^{(2)}&=&g_2^4 \left(\frac{9 g_1^2 }{10}+\frac{35 g_2^2}{6}+12  g_3^2-\frac{3  y_t^2}{2}-\frac{1}{2}  {\rm Tr}(Y^\dagger Y)\right),\nonumber 
\\
 \beta_{g_3^2}^{(2)}&=&g_3^4 \left(\frac{11 g_1^2}{10}+\frac{9g_2^2}{2}-\frac{40 g_3^2}{3}-2 y_t^2- y^2\right),\nonumber  \\
 \beta_{y_t^2}^{(2)}&=&+y_t^2 \bigg[ 6\lambda_H^2 -\frac{23 g_2^4}{4}+
\nonumber \\ && \hspace{-1.5cm}
y_t^2 \left( -12 y_t^2 -12 \lambda_H 
+36 g_3^2+\frac{225 g_2^2}{16}+\frac{393 g_1^2}{80} -\frac94  {\rm Tr}(Y^\dagger Y) \right)
\nonumber \\ && \hspace{-1cm}
+ \frac{1187 g_1^4}{600}+9 g_3^2 g_2^2+ 
   \frac{19}{15} g_3^2 g_1^2-\frac{9}{20} g_2^2 g_1^2  -\frac{932 g_3^4}{9}\nonumber \\ && \hspace{-1cm}
    +\left(\frac{3 g_1^2}{8}+\frac{15 g_2^2}{8} \right){\rm Tr}(Y^\dagger Y) -\frac94{\rm Tr}((Y^\dagger Y)^2)  +\frac{\lambda_{HA}^2}{2} \bigg],    \nonumber
   \\
\beta_{\lambda_H}^{(2)}&=&\lambda_H^2 \left[54 \left(g_2^2+\frac{g_1^2}{5}\right) \right. \nonumber \\ &&  -156\lambda_H -72 y_t^2 -24{\rm Tr}(Y^\dagger Y)\bigg] 
\nonumber \\ && \hspace{-1cm}
+\lambda_H y_t^2 \left( 40 g_3^2
+\frac{45 g_2^2}{4}+\frac{17 g_1^2}{4}-\frac32 y_t^2\right)
\nonumber \\ &&\hspace{-1cm}  +\lambda_H \bigg[\frac{1887 g_1^4}{400} -\frac{73 g_2^4}{16}
+\frac{117 g_2^2 g_1^2}{40} \nonumber \\ && \hspace{-1cm} 
+\left(\frac{3 g_1^2}{4}+\frac{15 g_2^2}{4} \right){\rm Tr}(Y^\dagger Y)\nonumber \\ && \hspace{-1cm}
-\frac{{\rm Tr}((Y^\dagger Y)^2}{2}) +20 g_3^2 y^2-\frac{9 y^4}{2}-5 \lambda_{HA}^2\bigg]
\nonumber \\ &&\hspace{-1cm}  +y_t^4 \left( 15 y_t^2-16g_3^2-\frac{4 g_1^2}{5}\right)
\nonumber \\ &&\hspace{-1cm} +y_t^2 \left(\frac{63 g_2^2g_1^2}{20} -\frac{9 g_2^4}{8}-\frac{171 g_1^4}{200}\right)
\nonumber \\ &&\hspace{-1cm}  +\frac{305 g_2^6}{32} -\frac{3411 g_1^6}{4000} -\frac{289 g_2^4 g_1^2}{160} -\frac{1677 g_2^2 g_1^4}{800} 
 \nonumber \\ &&\hspace{-1cm}  - \left(\frac{9 g_1^4}{200} + \frac{3 g_1^2 g_2^2}{20}+ \frac{3 g_2^4}{8}\right){\rm Tr}(Y^\dagger Y)+5{\rm Tr}((Y^\dagger Y)^3)
  \nonumber \\ &&\hspace{-1cm} 
  -3y^2\lambda_{HA}^2-2\lambda_{HA}^3.  \nonumber \\
 \nonumber
\eea 
The RGEs in the $\overline{\rm MS}$ scheme are gauge invariant as proved in~\cite{Buttazzo:2013uya}.

Next, we consider what happens in crossing  the threshold $M_A$: as discussed in \cite{RandjbarDaemi:2006gf,EliasMiro:2012ay,Salvio:2015cja} one has to take into account a scalar threshold effect: in the low energy effective theory below $M_A$  one has the effective Higgs quartic coupling 
\be \lambda= \lambda_H-\frac{\lambda_{HA}^2}{4 \lambda_A}. \label{lambda_eff}\ee
 In practice one should do the following: below $M_A$ the RGEs are  the ones  given above with $\beta_{\lambda_{HA}}$ and $\beta_{\lambda_{A}}$ removed, $\lambda_A$ and $\lambda_{HA}$ set to zero and  $\lambda_H$ replaced by $\lambda$. Above $M_A$ one should include $\lambda_A$, $\lambda_{HA}$, $\beta_{\lambda_{A}}$ and $\beta_{\lambda_{HA}}$ and find $\lambda_H$ using the full RGEs and the boundary condition in (\ref{lambda_eff}) at $\bar \mu= M_A$.

As far as the new fermions are concerned, 
following  \cite{Casas:1999cd} we adopt the approximation in which the new Yukawa couplings run only above the corresponding mass thresholds; this can be technically implemented  by substituting $Y_{ij} \rightarrow Y_{ij} \theta(\bar{\mu}-M_j)$ and $y	\rightarrow y \theta(\bar{\mu} - M_q)$ on the right-hand side of the RGEs.

\section{Stability analysis}\label{Stability analysis}

Since we use the 1-loop   RGEs of the non-SM parameters, we approximate the  effective potential $V_{\rm eff}$ of the model with its {\it RG-improved tree-level potential}: we substitute to the bare couplings in the classical potential the corresponding running parameters. 
 
 \begin{figure}[t]
\begin{center}
 \includegraphics[scale=0.6]{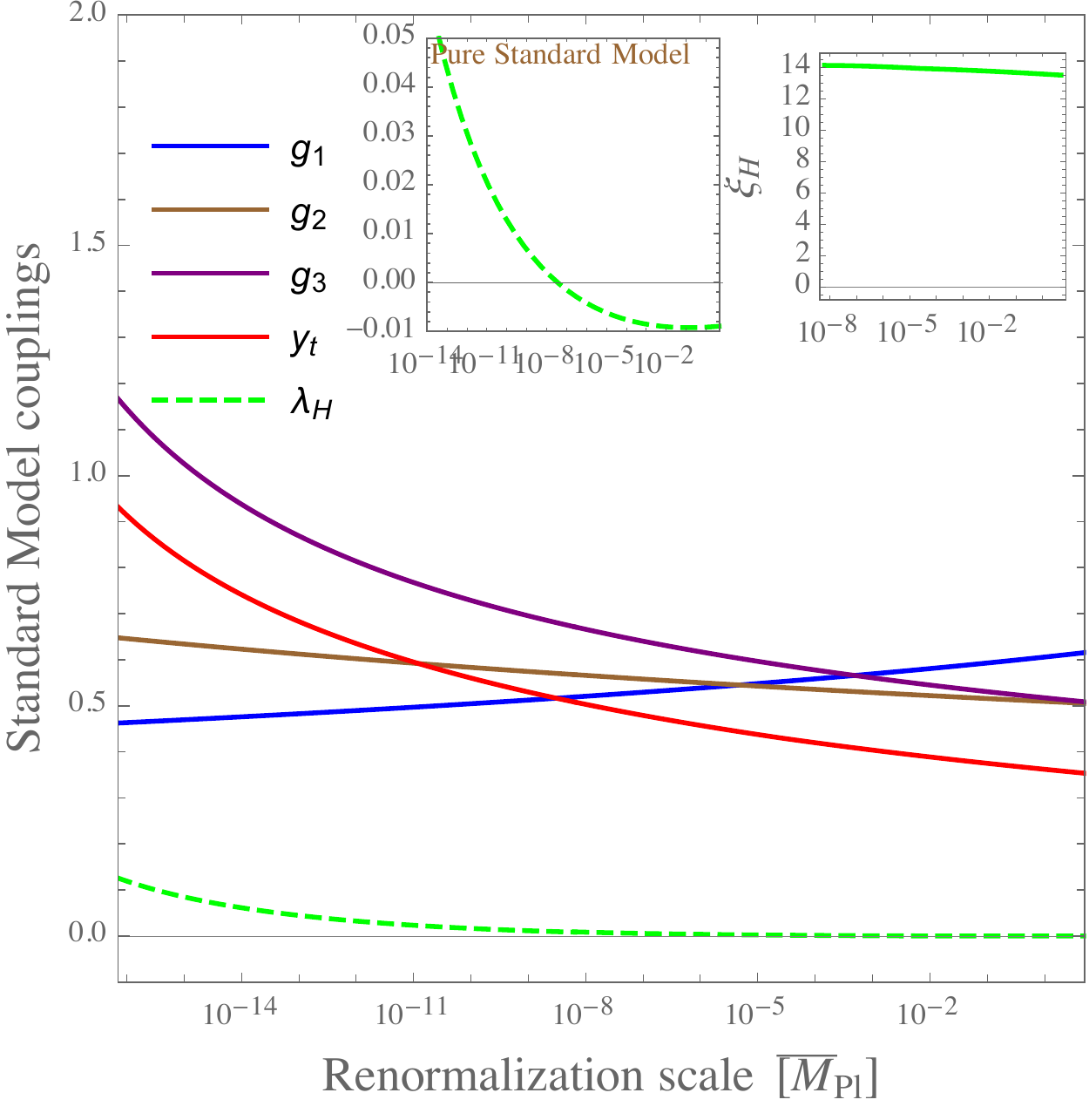}  
 \end{center}
   \caption{\em RG evolution of the relevant SM parameters  close to criticality ($\lambda_H$ is nearly zero at the Planck scale). The values of the parameters are the following: $M_1 = 10^{11}\,{\rm GeV}$, $M_2 = 6.5 \times 10^{13}\,{\rm GeV}$, $M_3 > \bp$, $z=0$, $\xi =1$,
    $f_a \simeq 2.5 \times 10^{10}\,{\rm GeV}$, $\lambda_{HA}(M_A)\simeq 0.016$, $\lambda_A(M_A) \simeq 0.1$, $y(M_A)\simeq  0.09$.  The inset on the left shows the running of $\lambda_H$ in the pure SM for the same values of the SM parameters.}
\label{SMcrit}
\end{figure}

Let us find the conditions that ensure the absolute stability  of the vacuum $\langle H_0\rangle = v$ and $\langle A\rangle =f_a$. We offer a more detailed treatment than the one in      
\cite{EliasMiro:2012ay} although we will agree with their conclusions.  For $v\ll f_a$, which is amply fulfilled thanks to (\ref{bound-f}), the conditions  are 
\begin{description}
\item[I.] $\lambda_H >0$ and $\lambda_A >0$
\item[II.]   $4\lambda_H \lambda_A -\lambda_{HA}^2 >0$
\end{description}
The origin of Condition {\bf I} is obvious. Notice, however, that once $\lambda_H>0$ and $\lambda_{HA}^2 <4\lambda_H\lambda_A $ are fulfilled then $\lambda_A>0$ is fulfilled too. The origin of Condition {\bf II} is provided in~\ref{Derivation of the stability conditions}.

An important remark is in order now. Suppose that, taking into account the dependence of the couplings on $\bar \mu$, one finds that Conditions {\bf I} and {\bf II} are violated at some energy $\bar \mu=\mu^*$. Can we really conclude that there is an instability? The answer to this question is ``yes" only  if $\mu^*$  is close enough to the field configurations at which the potential is lower than its value at the EW vacuum (henceforth the instability configurations); indeed, if this is not the case this instability would be outside the range of validity of the RG-improved tree-level potential. For this reason it is interesting to find the instability configurations. This is done in~\ref{Instability configurations}.

\begin{figure}[t]
\begin{center}
 \includegraphics[scale=0.6]{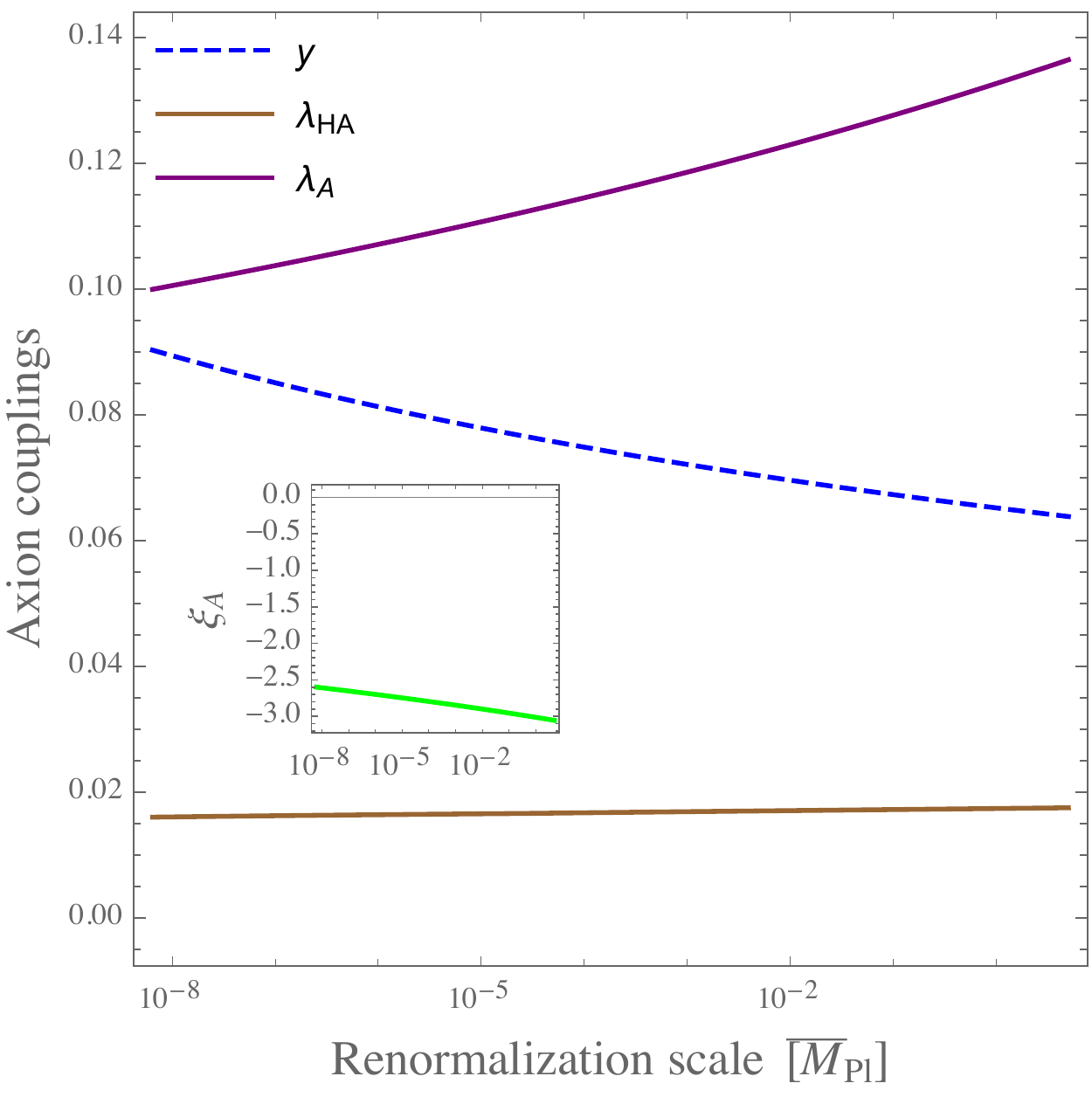}  
 \end{center}
   \caption{\em RG evolution of   the couplings of the axion sector   close to criticality ($\lambda_H$ is nearly zero at the Planck scale). The values of the parameters are as in Fig.~\ref{SMcrit}.}
\label{SMcrit2}
\end{figure}

\begin{figure}[t]
\begin{center}
 \includegraphics[scale=0.6]{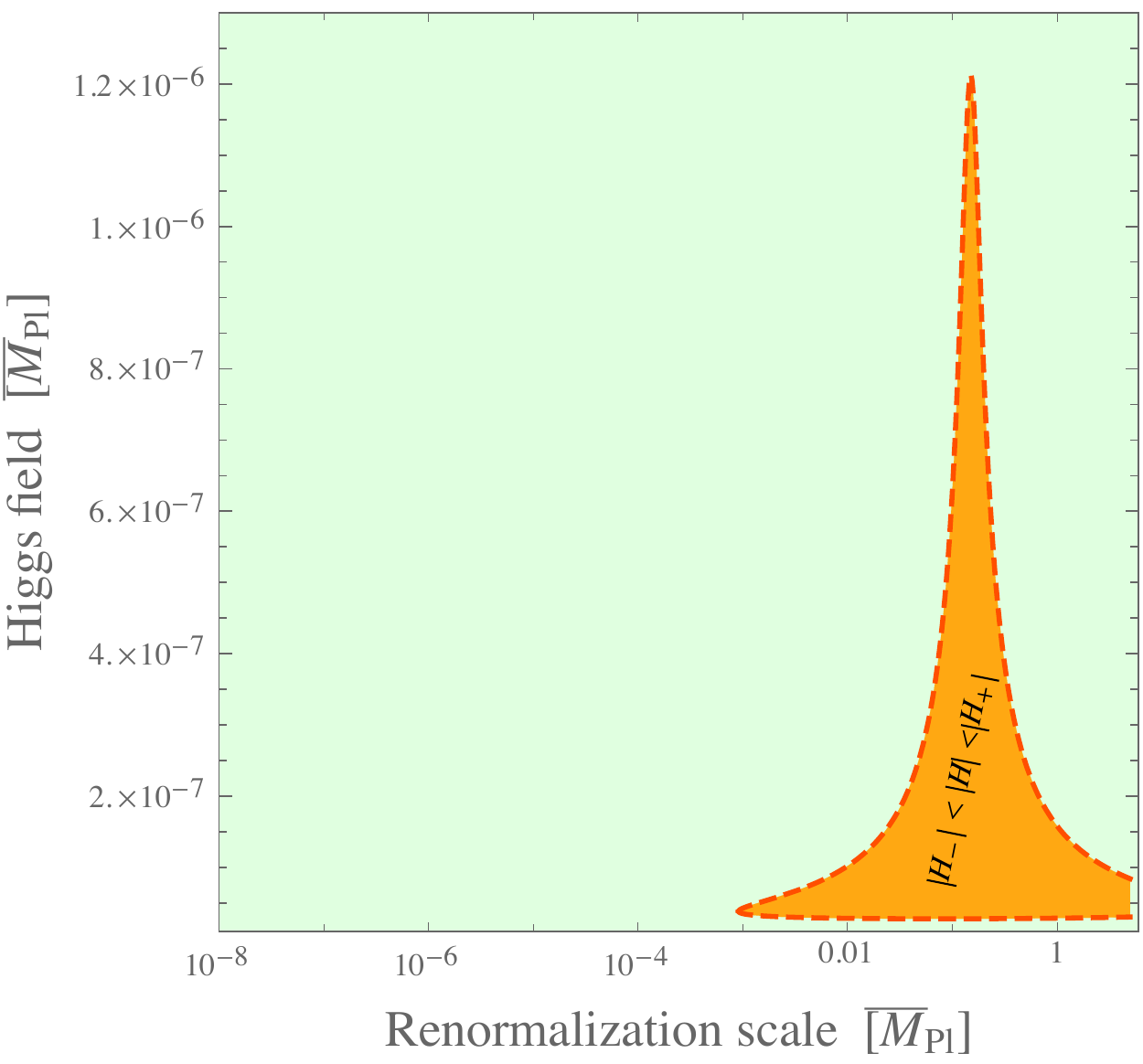}  
 \end{center}
   \caption{\em Configuration space in~(\ref{interval-instability}) and~(\ref{formHpm}) for the parameters set in Fig.~\ref{SMcrit}. This space opens up only at $\bar\mu \gg |H_{\pm}|$ so no instability configurations are found.}
\label{putative}
\end{figure}

In Fig.~\ref{SMcrit}  an example of  the running of the SM parameters close to criticality (and compatible with absolute stability) is provided (the example is specified in the caption). In that plot the threshold effect in (\ref{lambda_eff}) has been taken into account, but the jump of $\lambda_H$ cannot be appreciated in the plot  because a $\lambda_{HA} \ll \lambda_A$ has been chosen there. In the inset on the left the corresponding running of $\lambda_H$ in the pure SM is shown; note that in the pure SM one does not achieve absolute stability for the SM parameters used (which involve the current central value for $M_t$). Indeed, one of the main {\it testable} differences between the model discussed here and the pure SM is the possibility to realize CHI with the central value of $M_t$. In Fig.~\ref{SMcrit2}   the corresponding running of the couplings in the axion sector is shown. No pathologies (such as Landau poles) appear below the Planck scale and Condition {\bf I} for stability is satisfied. In Fig.~\ref{putative} the corresponding instability configurations for Condition {\bf II} (the configuration space defined in~(\ref{interval-instability}) and~(\ref{formHpm})) is shown: this space opens up only at $\bar\mu \gg |H_{\pm}|$ meaning that we do not encounter any instability of the EW vacuum (see the discussion in the previous paragraph).  This is not in contradiction with Figs.~1 and~2 of~\cite{Salvio:2015cja} because the region marked as ``$\lambda_{HA}^2(\bar{\mu}) < 4\lambda_H(\bar{\mu}) \lambda_A(\bar{\mu})$" there corresponds to having that condition satisfied for {\it all} $\bar\mu$ up to the Planck scale; as explained in the paragraph above this is only a sufficient  condition for absolute stability (not a necessary one). Note that for those parameter values neutrino data are reproduced, the axion accounts for the full DM abundance (see (\ref{bound-f-DM})) and leptogenesis can provide the observed matter-antimatter asymmetry. 

It is interesting to note that the criticality condition is not only achieved by a single point in the parameter space, but rather by living on a critical higher-dimensional surface. As a side comment, this opens up the intriguing possibility of lowering the masses of the new states close to a scale compatible with the Higgs naturalness and reachable at the LHC and/or future colliders.

In the next section it will be shown that a successful inflation can also be achieved with the Higgs close to criticality.

\section{Higgs inflation and criticality}\label{crit}

The possibility that we study in this article is that inflation is  triggered by the Higgs and in particular when one is very close to criticality.  During inflation the field values are very high (around the Planck scale), therefore,  the VEVs can be neglected as done in~\cite{Ballesteros:2016xej}. Indeed, even the highest VEV, $f_a$, is always  many orders of magnitude below the Planck scale thanks to~(\ref{bound-f-DM}). In this high-field regime, HI occurs when the quantities
 \be 
 \kappa_H \equiv\frac12 \lambda_{HA}\xi_H-\lambda_H\xi_A,
  \qquad \kappa_A\equiv \frac12 \lambda_{HA}\xi_A-\lambda_A \xi_H \label{SMASHcond}
 \ee
 satisfy 
 \be \{\kappa_H>0, \, \kappa_A < 0\}. \label{SMASHcond2} \ee
   Indeed, it turns out that the Higgs direction is a valley, while $|A|$ is a ridge of the   tree-level potential when~(\ref{SMASHcond2}) holds~\cite{Ballesteros:2016xej}.  Such situation occurs because of an interplay between the dimensionless parameters: the non-minimal couplings and the quartic couplings (as clear from (\ref{SMASHcond})-(\ref{SMASHcond2})). Therefore, in case~(\ref{SMASHcond2}) the inflation along the $|A|$-direction and the multifield inflation (in which both $|A|$ and $|H|$ are active) does not occur.

 In HI,  the action that involves the field $A$ can be neglected (because of the argument in the previous paragraph) and the term in the action that depends on the metric and the Higgs field  {\it only} (the scalar-tensor part) is 
\begin{equation} S_{\rm st} = \int d^4x\sqrt{-g}\left[|\partial H|^2-V_H-\left(\frac{\bp^2 }{2}+\xi_H |H|^2\right)\mathscr{R}\right], \label{Jordan-frame}\end{equation}
where  
 $V_H=\lambda_H |H|^4$ is the classical Higgs potential and we have ignored the EW scale $v$, which is  completely negligible compared to the inflationary scales (that will be discussed in this section and the next one). We assume a sizable non-minimal coupling, $\xi_H>1$, because this is what inflation leads to as we will see.

We start by using the classical approximation, later we will also include quantum corrections. The $\xi_H |H|^2 \mathscr{R}$ term can be removed through a {\it conformal} transformation (a.k.a. Weyl transformation):
\begin{equation} g_{\mu \nu}\rightarrow   \Omega_H^{-2}  g_{\mu \nu}, \quad \Omega_H^2= 1+\frac{2\xi_H |H|^2}{\bp^2}, \label{transformation}\end{equation}
which, as we will see below, redefines the kinetic term and the potential of the Higgs field.
The original frame, where the Lagrangian has the form in Eq.~(\ref{Jordan-frame}), is known as the Jordan frame, while the one where gravity is canonically normalized (after the transformation above) is called the Einstein frame. In the unitary gauge, where the only scalar field is   $\phi \equiv \sqrt{2|H|^2}$,  we have (after  having performed the conformal transformation)
\begin{equation} S_{\rm st} = \int d^4x\sqrt{-g}\left[K_H \frac{(\partial \phi)^2}{2}-\frac{V_H}{\Omega_H^4}-\frac{\bp^2 }{2}\mathscr{R}\right], \label{Sst}\end{equation}
and 
\be K_H =  \Omega_H^{-4} \left[\Omega_H^2 +\frac{3\bp^2}{2}  \left(\frac{d\Omega_H^2}{d\phi}\right)^2 \right]. \label{K1} \ee

The non-canonical Higgs kinetic term can be made canonical through the  field redefinition $\phi=\phi(\phi')$ given by
\begin{equation} \frac{d\phi'}{d\phi}= \Omega_H^{-2} \sqrt{\Omega_H^2 +\frac{3\bp^2}{2}  \left(\frac{d\Omega_H^2}{d\phi}\right)^2 } ,\label{phi'}\end{equation}
with the conventional condition $\phi(\phi'=0)=0$. Note that $\phi(\phi')$ is invertible because Eq.~(\ref{phi'}) tells us $d\phi'/d\phi> 0$. 
Thus, one can extract the function $\phi(\phi')$ by inverting the function $\phi'(\phi)$ defined above. We will refer to $\phi'$ as the canonically normalized Higgs field.
Note that $\phi'$ feels a potential 
\begin{equation} U_H\equiv \frac{V_H}{\Omega_H^4}=\frac{\lambda_H\phi(\phi')^4}{4(1+\xi_H\phi(\phi')^2/\bp^2)^2}\label{UH} . \end{equation}

Let us now recall how  inflation emerges in this context in the slow-roll approximation. From Eqs.~(\ref{phi'}) and~(\ref{UH}) it follows  that $U_H$ is exponentially flat when $\phi' \gg \bp$~\cite{Bezrukov:2007ep}, which is a key property to have inflation. Indeed, for such high field values the slow-roll parameters
\be \epsilon_H \equiv\frac{\bp^2}{2} \left(\frac{1}{U_H}\frac{dU_H}{d\phi'}\right)^2, \quad \eta_H \equiv \frac{\bp^2}{U_H} \frac{d^2U_H}{d\phi'^2} \label{epsilon-def}\ee
are guaranteed to be small. Therefore, the region in field configurations where $\phi' \gtrsim \bp$ (or equivalently \cite{Bezrukov:2007ep} $\phi \gtrsim \bp/\sqrt{\xi_H}$) corresponds to inflation. 

The parameter $\xi_H$ can be fixed by requiring that the predicted curvature power spectrum equals the observed value, Eq.~(\ref{PRobserved}), for a field value $\phi'=\phi'_{\rm b}$ corresponding to an appropriate number of e-folds \cite{Bezrukov:2008ut}:

\begin{equation}N=\int_{\phi'_e}^{\phi'_{\rm b}}\frac{U_H}{\bp^2}\left(\frac{dU_H}{d\phi'}\right)^{-1}
d\phi',
\label{e-folds}\end{equation}
where $\phi'_e$ is the field value at the end of inflation, computed by requiring 
\be \epsilon_H(\phi'_e) \simeq 1. \label{inflation-end}\ee
In the slow-roll approximation (used here) such constraint can be imposed by using the standard formula
\begin{equation}P_{R}(k)= \frac{U_H/ \epsilon_H}{24\pi^2 \bp^4}.\label{PRsr} \end{equation}
 For $N\sim 60$, this procedure leads to a very large $\xi_H$ at the classical level. However, the need of a very large $\xi_H$ can be avoided when quantum corrections are included \cite{Hamada:2014iga,Bezrukov:2014bra,Hamada:2014wna}, as we will see below.

We can also compute the scalar spectral index $n_s$ and  the tensor-to-scalar ratio $r$: in the slow-roll approximation   the  formul$\ae$  are  $r =16\epsilon_H$ and 
$n_s=1-6\epsilon_H +2\eta_H$.
These parameters are important as they are constrained by observations (as we have seen in Sec.~\ref{Observational bounds}).
 
 We now discuss the quantum corrections to the Higgs potential. We want to include both the large-$\xi_H$ inflationary scenario of \cite{Bezrukov:2007ep} and the CHI proposed in \cite{Hamada:2014iga,Bezrukov:2014bra,Hamada:2014wna}. The latter case permits a drastic decrease of the value of $\xi_H$ with respect to the classical result. This indicates that we cannot rely on large-$\xi_H$ approximations to analyze this case. Thus, we do not use such approximations here. However, we do assume in the following that $\xi_H > 1$ as this is present both in the original formulation of HI  and in CHI. 

Note that Eqs.~(\ref{Jordan-frame}), (\ref{transformation}), (\ref{K1}) and (\ref{phi'}) also hold if $\xi_H$ is field-dependent, as dictated by quantum corrections~\cite{Ezquiaga:2017fvi}.
A second step we should do now is the computation of the  effective potential. In defining the quantum theory there are well-known ambiguities~\cite{Bezrukov:2009db,Bezrukov:2009-2, Bezrukov:2014bra,Bezrukov:2014ipa,Bezrukov:2017dyv}. We follow here Ref.~\cite{Bezrukov:2014bra} and choose to compute the loop corrections to the effective potential - a.k.a. Coleman-Weinberg potential - in the Einstein frame (after having performed the conformal transformation  (\ref{transformation})). This choice is convenient because we can then use the standard formul$\ae$ to compute the primordial quantum fluctuations, which assume minimal couplings to gravity. The effective potential is also RG-improved by using the RGEs.

 Such prescription to compute the quantum effects is known as Prescription I and it leads to the following renormalization group scale 
\be \bar\mu(\phi) =  \frac{\phi/\kappa}{\sqrt{1+\xi_H \phi^2/\bp^2}}, \label{mubar}\ee
where $\kappa$ is an order one  factor.  

 In a previous work~\cite{Bezrukov:2014ipa} ``threshold corrections" at the scale $\bp/\xi_H$ for the RG flow have been considered. We regard such corrections as different ways of quantizing the theory, which are non-minimal as they require further parameters. Furthermore, in some UV modifications of general relativity the above-mentioned corrections vanish~\cite{Salvio:2014soa,Salvio:2017qkx,Salvio:2018crh}. For these reasons we will not consider them in this article.

Furthermore,  we will use the RG-improved potential neglecting the loop corrections: this means that we will take as effective potential the one in Eq.~(\ref{UH}) with the constants $\lambda_H$ and $\xi_H$ replaced by the corresponding running parameters. There are good reasons to use this approximation. 
Indeed, taking into account the loop corrections to the potential would only be more precise if supplemented by the loop corrections to the kinetic term of the inflaton; such corrections have not been included in   HI and are expected to be comparable to the loop corrections to the potential for moderate values of $\xi_H$, unlike what happens for large $\xi_H$ \cite{Bezrukov:2009-2}: the large value of $\xi_H$ allowed~\cite{Bezrukov:2009-2} to show that the corrections to the kinetic term are negligible, but the smaller value of $\xi_H$ of critical HI does not permit to trust this approximation anymore. 
Another reason to employ the RG-improved potential is its gauge independence, which is not shared by the Coleman-Weinberg effective potential. Therefore, the use of the RG-improved potential allows us to obtain  a more transparent physical interpretation.

Given that we use this approximation we can also compute the RGEs in the Jordan frame. Let us see why. In an exact computation we should also compute the RGEs in the Einstein frame (just like the Coleman-Weinberg potential) but the approximation in which the RGEs are computed in the Jordan frame is a good approximation because the error one is doing is of order of the Weyl anomaly, which is suppressed by $1/(4\pi)^2$~\cite{Duff:1977ay} and we are not including anyway the Coleman-Weinberg corrections to the potential which are of the same order.

\begin{figure}[t]
\begin{center}
 \includegraphics[scale=0.6]{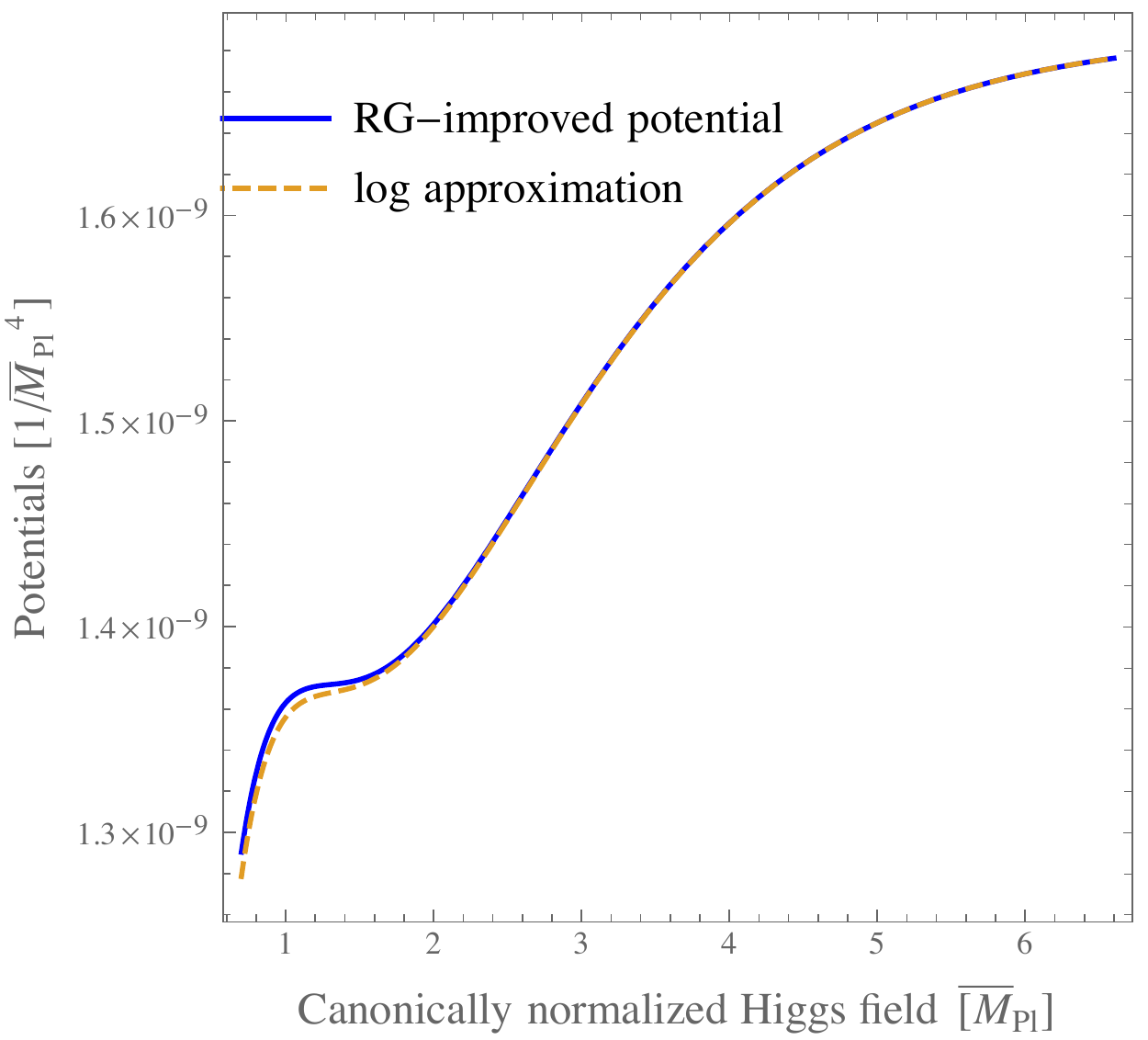}  
 \end{center}
   \caption{\em RG-improved potential and its approximation (log-approximation) based on the expansions in Eq.~(\ref{log-expansion}) for the parameters set in Fig.~\ref{SMcrit} and for $\kappa\simeq 1.8$. }
\label{log-approximation}
\end{figure}

Moreover, in computing the inflationary potential a further approximation can be done. One can approximate the running couplings $\lambda_H$ and $\xi_H$ by expanding them around the minimum of $\lambda_H$ (henceforth $\lambda_0$), which typically occurs around the Planck scale, as follows
\be \lambda_H(\bar\mu) \simeq \lambda_0 + b_\lambda \ln^2(\bar\mu/\mu_0), \qquad  \xi_H(\bar\mu) \simeq \xi_0 + b_\xi \ln(\bar\mu/\mu_0),\label{log-expansion}\ee
where  $\mu_0$ is the value of $\bar\mu$ where this minimum occurs, and $\xi_0 \equiv \xi_H(\mu_0)$. The parameters $b_\lambda$ and $b_\xi$  are related to the $\beta$-functions as follows
\be b_\lambda = \left. \bar\mu \frac{d\beta_{\lambda_H}}{d\bar\mu} \right|_{\bar\mu=\mu_0}, \qquad b_\xi = \left.2\beta_{\xi_H} \right|_{\bar\mu=\mu_0}.\ee
and can be computed once the RGEs are solved. Then, one can approximate the potential by inserting these expansions inside~(\ref{UH}). Such approximation (which we will call the ``log approximation") works very well (see Fig.~\ref{log-approximation}) and we will therefore use it from now on.

Now, Eqs.~(\ref{epsilon-def}), (\ref{e-folds}) and (\ref{PRsr})  are still valid as long as one is in the slow-roll regime, but one should now interpret $U$ as the effective potential, not just as the classical  potential.

 
 The inflationary observables predicted by the model analyzed here are in agreement with the most recent observational bounds~\cite{Planck2018} (see  for instance Eqs~(\ref{nsr}) and~(\ref{PRobserved})): e.g. for the parameters used in Fig.~\ref{SMcrit} we have
 \be n_s \simeq 0.965, \qquad r = 0.0472, \qquad P_R=2.12\times 10^{-9} \ee
 and a number of e-folds equal to about 55. 
 
 Note that for the values of the parameters used in Fig.~\ref{SMcrit} one has $\{\kappa_H>0, \kappa_A < 0\}$ (see also Fig.~\ref{SMcrit2}) and, therefore, as discussed at the beginning of this section, in that setup the inflation along the $|A|$-direction and the multifield inflation (in which both $|A|$ and $|H|$ are active) does not occur.


Furthermore, in Ref.~\cite{Salvio:2015kka} it was shown that CHI features a robust inflationary attractor in the SM. The same conclusion holds here because the results of the analysis in Ref.~\cite{Salvio:2015kka} were based only on the qualitative features of the inflationary potential, which are the same in the model studied here. Moreover, for $\{\kappa_H>0, \kappa_A < 0\}$, which has been realized in this paper, the other directions in the scalar field space are {\it not} inflationary attractors~\cite{Ballesteros:2016xej}.

\section{Validity of the effective theory}\label{Validity of the effective theory}

We have already commented that CHI leads to an  increasing of the cutoff of the effective theory on flat spacetime compared to the ordinary HI case. Let us generalize the discussion now to include the non-trivial background fields characteristics of inflation. Ref.~\cite{Bezrukov:2010jz}  showed that the cutoff of the theory  can be studied by dividing the range of  the background Higgs field $\bar\phi$  into three regimes\footnote{Note that $\bar\phi$ is the background value of $\phi$ not of $\phi'$.}. We use the results of Ref.~\cite{Bezrukov:2010jz}  in the following and further extend them.
\begin{itemize}
\item $\bar\phi \ll \bp/\xi_H$. In this small field regime the cutoff of the theory is identified as the coefficients of the dimension-$n$ operators $\delta \phi'^n$ (for $n>4$), where $\delta \phi'$ is the fluctuation of $\phi'$ around its background value $\bar\phi'$. The cutoff obtained in this way reads
\be \Lambda_{(n)} = \frac{\bp}{\xi_H} \lambda_H^{-1/(n-4)}, \ee
where $n$ acquires even values. This is the flat spacetime result. Given that in CHI $\lambda_H$ is very small the smallest value of the cutoff is reached in the limit $n\to \infty$. However, for moderate values of $n$ the cutoff $\Lambda_{(n)}$ is much bigger in CHI than in the ordinary HI case thanks to the smallness of $\lambda_H$. In Fig.~\ref{cutoffs1} we show this tower of cutoffs (varying $n$) as a function of the canonically renormalized Higgs field $\phi'$  and compare them with the inflationary scale, defined as $U_H^{1/4}$. In the plot we also take into account the running of the couplings. One finds that the inflationary scale is always much smaller than the cutoff.
\item  $\bp/\xi_H\ll \bar\phi \ll  \bp/\sqrt{\xi_H}$. By following a procedure similar to the one used in the small field regime,   the cutoff in this intermediate range  is instead
\be \Lambda_{(n)} = \frac{\bar\phi^2 \xi_H}{\bp} \left(\frac{\xi_H^6\bar\phi^6}{\lambda_H \bp^6}\right)^{1/(n-4)} \ee
and again in CHI the smallest value of the cutoff is obtained in the limit $n\to \infty$ and for moderate $n>4$ the CHI features a much larger cutoff than the large-$\xi_H$ HI. In this field range the tower of cutoffs is shown in Fig.~\ref{cutoffs2}, taking into account the running of the couplings, and compared again with the inflationary scale. That plot shows that  the inflationary scale is always much smaller than the cutoff.

\item Finally, in the inflationary regime, $ \bar\phi \gg \bp/\sqrt{\xi_H}$, 
the cutoff is simply $\Lambda \sim \bp$, which coincides with the scale at which sizable  quantum gravity effects are expected to emerge. Fig.~\ref{log-approximation} shows that the inflationary scale is much smaller than the cutoff in this last regime too.
\begin{figure}[t]
\begin{center}
 \includegraphics[scale=0.6]{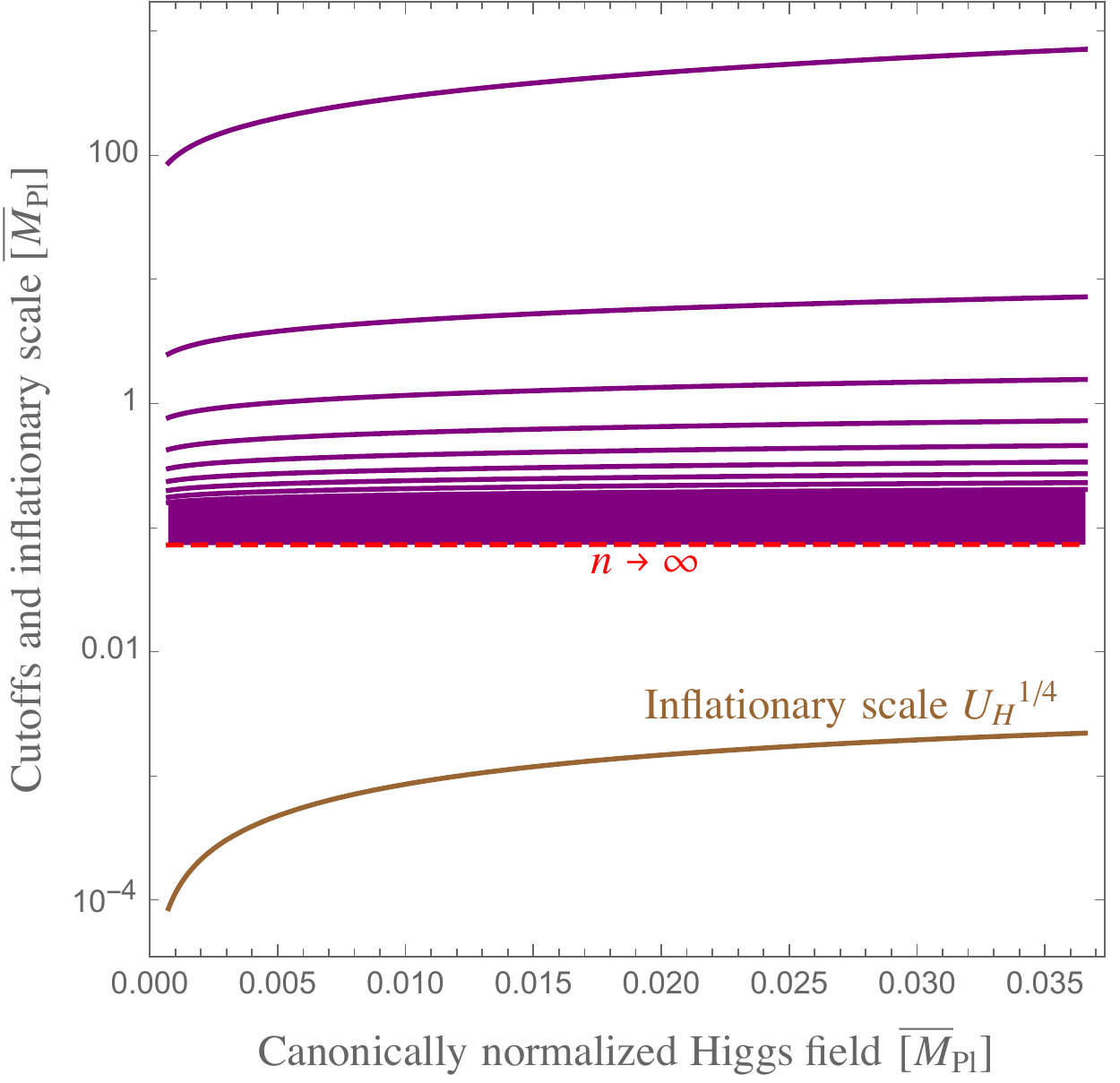}   
 \end{center}
   \caption{\em The cutoff of the theory obtained by reading the coefficients of the dimension-$n$ operators $\delta \phi'^n$ (for $n>4$ and varying $n$) is compared to the inflationary scale. The parameters are chosen as in Fig.~\ref{log-approximation}.   In this plot the small field regime ($\bar\phi \ll \bp/\xi_H$) is shown. }
\label{cutoffs1}
\end{figure}

\begin{figure}[t]
\begin{center}
   \includegraphics[scale=0.6]{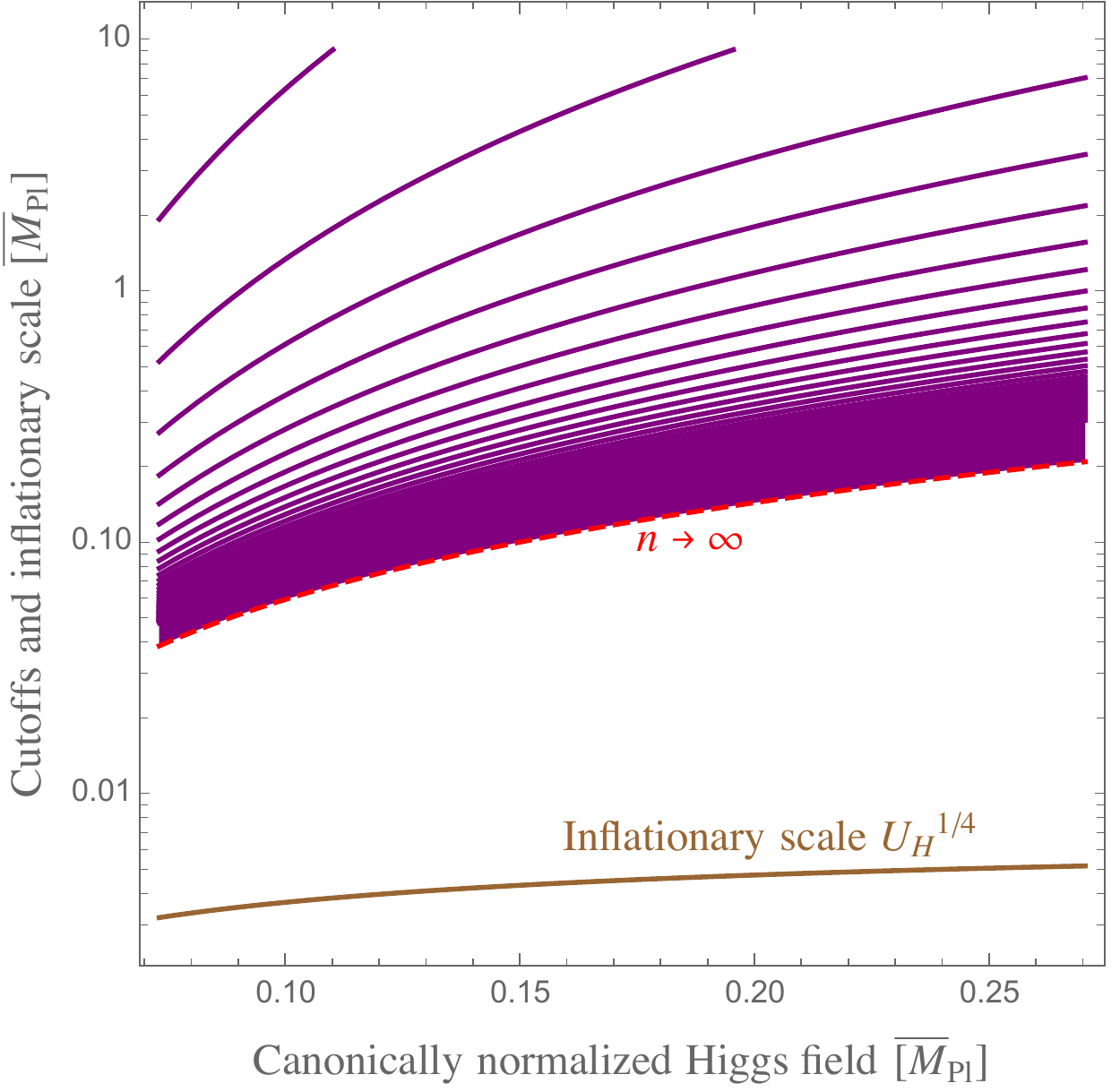}  
 \end{center}
   \caption{\em The same as in Fig.~\ref{cutoffs1}  but here  the intermediate field regime ($\bp/\xi_H\ll \bar\phi \ll  \bp/\sqrt{\xi_H}$) is shown.}
\label{cutoffs2}
\end{figure}
\end{itemize}
 
\section{Conclusions}\label{Conclusions}

\vspace{-0.1cm}

  In this article it was found that CHI can be implemented in a well-motivated extension of the SM, which explains with few extra degrees of freedom neutrino oscillations (through three right-handed neutrinos), DM (identified with the axion), baryon asymmetry (through leptogenesis) and the strong CP problem (thanks to the PQ symmetry). Furthermore, all the above-mentioned features can be there together with a stable EW vacuum. The fact that CHI can be implemented in this context is non-trivial: indeed, to establish this result one needs to carefully take into account the stability conditions in the presence of RG-improved parameters.

 CHI inflation has the advantage of 1) being free from a large non-minimal coupling $\xi_H$ and from a consequent scale of violation of perturbative unitarity much below the Planck scale, 2) enjoying a robust inflationary attractor and 3) allowing for an efficient reheating thanks to the sizable couplings between the Higgs and other SM particles.

 Moreover, in the proposed model, unlike in  the SM~\cite{Bezrukov:2014bra,Masina:2018ejw}, CHI can be realized without any tension with the observed quantities, such as the top mass or the strong fine-structure constant.

\vspace{-0.1cm}

\section*{Acknowledgments} 
\vspace{-0.1cm} 

\noindent   I thank F.~L.~Bezrukov, I.~Masina and R.~Torre   
  for useful discussions. This work was supported by the grant 669668 -- NEO-NAT -- ERC-AdG-2014.

\vspace{-0.cm} 


\appendix

\section{\\ Derivation of the stability conditions}\label{Derivation of the stability conditions}

As already stated in the main text, the only stability condition that requires explanation is Condition {\bf II} (see Sec.~\ref{Stability analysis}). This condition can be derived as follows. 

Assume $\lambda_H> 0$ and $\lambda_A >  0$, which is anyhow required by the stability, and then define 
\be u\equiv \sqrt{\lambda_H}(|H|^2-v^2),\qquad  w\equiv \sqrt{\lambda_A} (|A|^2-f_a^2),\ee
 to write the potential in a more compact form:
\be V = u^2 + w^2 + \frac{\lambda_{HA}}{\sqrt{\lambda_H\lambda_A}} \, uw .\ee 
Note that 
\be u \in [-\lambda_H v^2, \infty], \qquad w \in [-\lambda_A f_a^2, \infty]. \label{ranges} \ee
Next  use polar coordinates, $u = r \cos\theta$, $w = r \sin\theta$, which give
\be V = r^2\left(1 +  \frac{\lambda_{HA}}{\sqrt{\lambda_H\lambda_A}} \sin\theta\cos\theta\right). \ee
The stability condition is that $V$ should not become negative for any field value, thus 
\be 1 +  \frac{\lambda_{HA}}{\sqrt{\lambda_H\lambda_A}} \sin\theta\cos\theta > 0. \label{StabCondII}\ee 
When $\sin\theta\cos\theta$ acquires its minimum ($\sin\theta\cos\theta = - 1/2$), inequality (\ref{StabCondII}) becomes 
\be \lambda_{HA} < +2 \sqrt{\lambda_H\lambda_A}\label{condIIb}\ee
and ensures that $V$ is non-negative for positive $\lambda_{HA}$.
When $\sin\theta\cos\theta$ acquires instead  its maximum ($\sin\theta\cos\theta = + 1/2$) this inequality becomes
\be \lambda_{HA} > -2 \sqrt{\lambda_H\lambda_A} \label{condIIa}\ee
and ensures that $V$ is non-negative for negative $\lambda_{HA}$. 
Putting together (\ref{condIIa}) and (\ref{condIIb}) gives
\be \lambda_{HA}^2 < 4 \lambda_H\lambda_A\label{condIItot}.\ee

\section{\\ Instability configurations}\label{Instability configurations}

In this appendix we find the field configurations at which the potential is lower than its value at the EW vacuum if Condition {\bf II} (see Sec.~\ref{Stability analysis}) is violated. We called these configurations the  ``instability configurations".

 Let us focus on the case $\lambda_{HA}>0$, since this case is the one that imposes the weaker constraint from stability~\cite{EliasMiro:2012ay}. From the discussion provided in~\ref{Derivation of the stability conditions} the most unstable direction is $\theta = \theta_0 \equiv  -\pi/4$ (at which $\sin\theta\cos\theta = - 1/2$) and the most unstable value of $r$ (henceforth $r_0$) is the maximal one compatible with the ranges in (\ref{ranges}). To compute $r_0$ note that the most unstable configuration has $w = w_0 \equiv -\sqrt{\lambda_A} f_a^2$ (we  assume $\lambda_H>0$ and $\lambda_A>0$ as required by Condition {\bf I}), which corresponds to 
\be A= A_0 \equiv  0. \ee From $r_0 \sin\theta_0 = w_0$ one then determines $r_0 = \sqrt{2\lambda_A} f_a^2$. Therefore, the most unstable configuration has $u = u_0 \equiv r_0 \cos\theta_0 = \sqrt{\lambda_A} f_a^2$, or, in terms of $|H_0|$, defined by $\sqrt{\lambda_H}(|H_0|^2 -v^2) = u_0$,
\be |H_0|^2 = v^2 +\sqrt{\frac{\lambda_A}{\lambda_H}} f_a^2. \ee
Moreover, the full instability configurations with $A =0$ are given by
\be |H_-|< |H|<|H_+| \label{interval-instability}\ee 
where 
\be |H_\pm|^2 = v^2 +\frac{f_a^2\lambda_{HA}}{2\lambda_H}\left(1\pm \sqrt{1-\frac{4\lambda_H\lambda_A}{\lambda_{HA}^2}}\right), \label{formHpm}\ee
which confirms the result in Ref.~\cite{EliasMiro:2012ay}. 
%

\end{document}